\documentclass[final
    ,final            
  ]
  {aipproc}

\layoutstyle{6x9}
\def\ie{{\it i.e.}}
\def\eg{{\it e.g.}}
\def\etc{{\it etc}}

\def\tev{\,{\ifmmode\mathrm {TeV}\else TeV\fi}}
\def\mev{\,{\ifmmode\mathrm {MeV}\else MeV\fi}}
\def\gev{\,{\ifmmode\mathrm {GeV}\else GeV\fi}}

\begin{document}

\title{Supersymmetry Without (Too Much) Prejudice}
\rightline{\vbox{\halign{&#\hfil\cr
&SLAC-PUB-13710\cr
}}}


\author{Thomas G.Rizzo}{
  address={SLAC National Accelerator Laboratory, 2575 Sand Hill Rd., 
Menlo Park, CA, 94025, USA}
}

\begin{abstract}
We have recently completed a detailed scan of the 19-dimensional parameter 
space of the phenomenological MSSM, \ie, the CP-conserving MSSM assuming Minimal 
Flavor Violation(MFV) with the first two sfermion generations degenerate. 
We found a large set of parameter space points that satisfied all of 
the existing experimental and theoretical constraints. This analysis 
allows us to examine the general features of the MSSM without reference 
to any particular SUSY breaking scenario or any other assumptions about physics 
at higher scales. This study opens up new possibilities for SUSY phenomenology 
both at colliders and in astrophysical observations.
\end{abstract}

\maketitle


Beyond the domains of the specific SUSY breaking scenarios (\eg, mSUGRA, AMSB, 
GMSB, \etc), how well do we really know the properties of the MSSM? Are there 
parameter space regions which lead to sparticle properties which produce new or different 
experimental signatures not previously encountered? Are atypical dark matter signals 
possible?  To begin to address these and related questions we{\cite {us,talks}} performed 
a pair of scans of a limited 19-dimensional 
subspace of the full 100+ parameter (R-parity conserving!) MSSM obtained by 
assuming a thermal relic neutralino LSP, CP conservation(\ie, only real soft breaking 
parameters), weak scale MFV   
and by taking the first 2 sfermion generations degenerate (flavor by flavor) with negligible 
Yukawa couplings. Note that we have made no assumptions about any high-scale 
physics in this analysis.  

To perform our scans, we need to choose suitable ranges for the soft-breaking SUSY 
parameters (evaluated at the TeV scale) and determine how the values of these parameters
are picked within these assumed ranges. We purposefully chose ranges that would lead to 
SUSY which is relatively easy to access kinematically at the LHC. For the analysis presented 
here, we actually made two independent parameter scans. In the first case, we 
randomly generated $10^7$ sets of parameters (\ie, models), 
assuming flat priors, \ie, we assumed that the parameter values are chosen 
{\it uniformly} throughout their allowed ranges.
For this flat prior scan we employed the following ranges for our 19 parameters:

\begin{eqnarray}
100 \gev \leq m_{\tilde f} \leq 1\tev \,, \nonumber\\
50\gev \leq |M_{1,2},\mu|\leq 1 \tev\,, \nonumber \\ 
100 \gev \leq M_3\leq 1 \tev\,, \nonumber \\ 
|A_{b,t,\tau}| \leq 1 \tev\,, \\
1 \leq \tan \beta \leq 50\,, \nonumber \\ 
43.5\gev \leq m_A \leq 1 \tev\,. \nonumber  
\end{eqnarray}

For our second scan, we randomly generated $2\times 10^6$ model points 
assuming log priors for the mass parameters with slightly modified allowed 
ranges. The goal of this second scan is to make contrasts and comparisons to 
the flat prior study in order to determine the dependence of the final model 
properties on the scan assumptions. For the case of our two priors we find 
that they yield qualitatively and semi-quantitatively very similar results. 

To obtain a valid model we required a very large number of constraints to be 
satisfied including those from flavor physics, precision measurements, the muon g-2, 
dark matter direct detection as well as those from Higgs and SUSY searches at both LEP 
and the Tevatron{\cite {us}} requiring detector simulation studies. The WMAP 
dark matter density constraint was also employed but only as an upper bound to allow for 
other dark matter to exist besides the LSP. Theoretical constraints, such as Higgs 
potential stability and the absence of color breaking minima were also imposed. 
For the flat(log) prior analysis this yielded $\sim 68.5(2.8)$k successful 
models whose properties we briefly discuss below{\cite {us}}. Once we have 
these model sets we can examine their properties to see how they differ from
the standard scenarios and discover if they lead to new experimental signatures  
not previously considered{\cite {talks}}. Note that our goal is not to find 
the {\bf best-fit } models here but to find successful ones that produce new 
signatures.

\begin{figure}
\centerline{
  \includegraphics[height=.27\textheight]{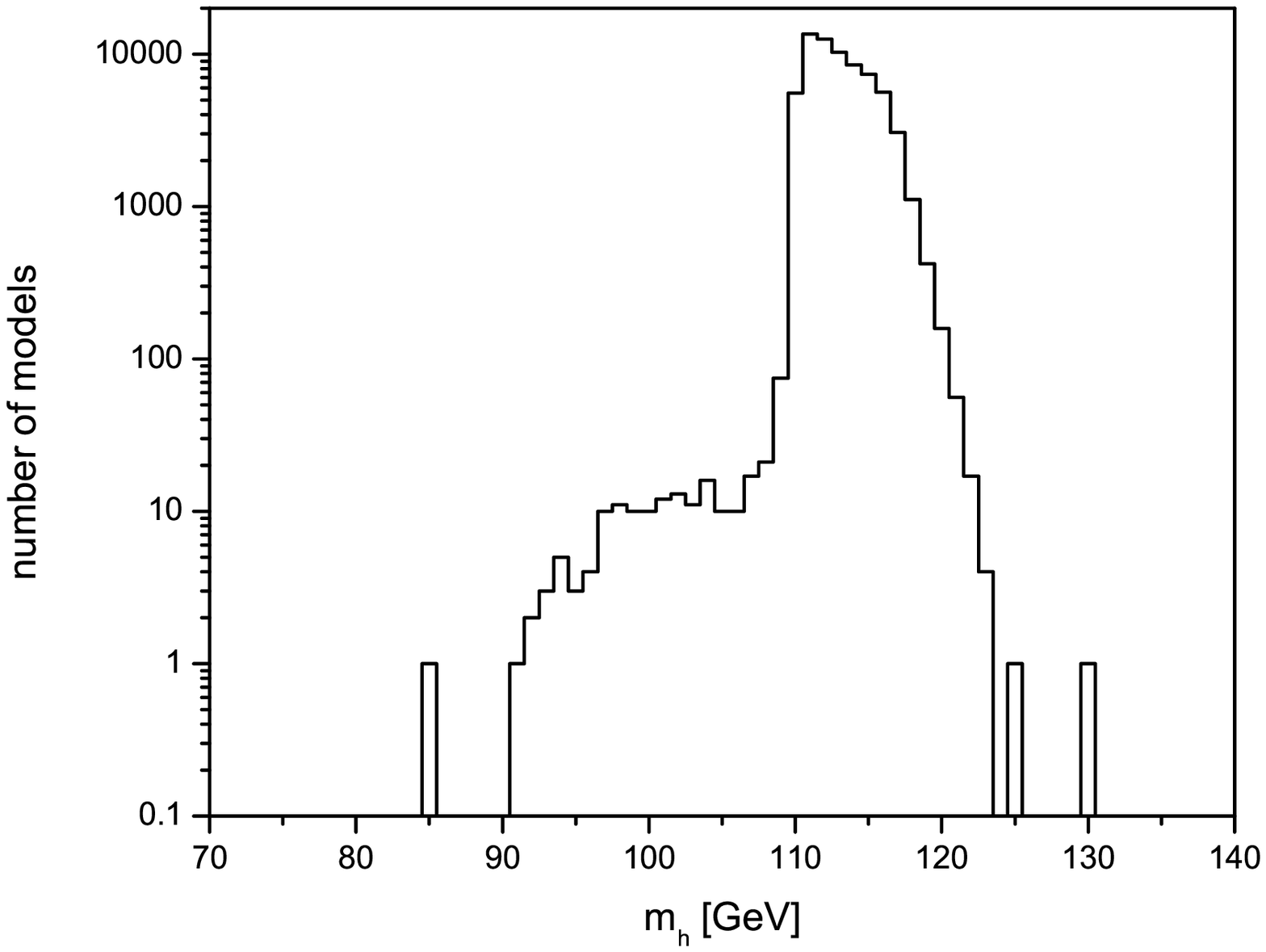}
\hspace*{-0.1cm}
  \includegraphics[height=.27\textheight]{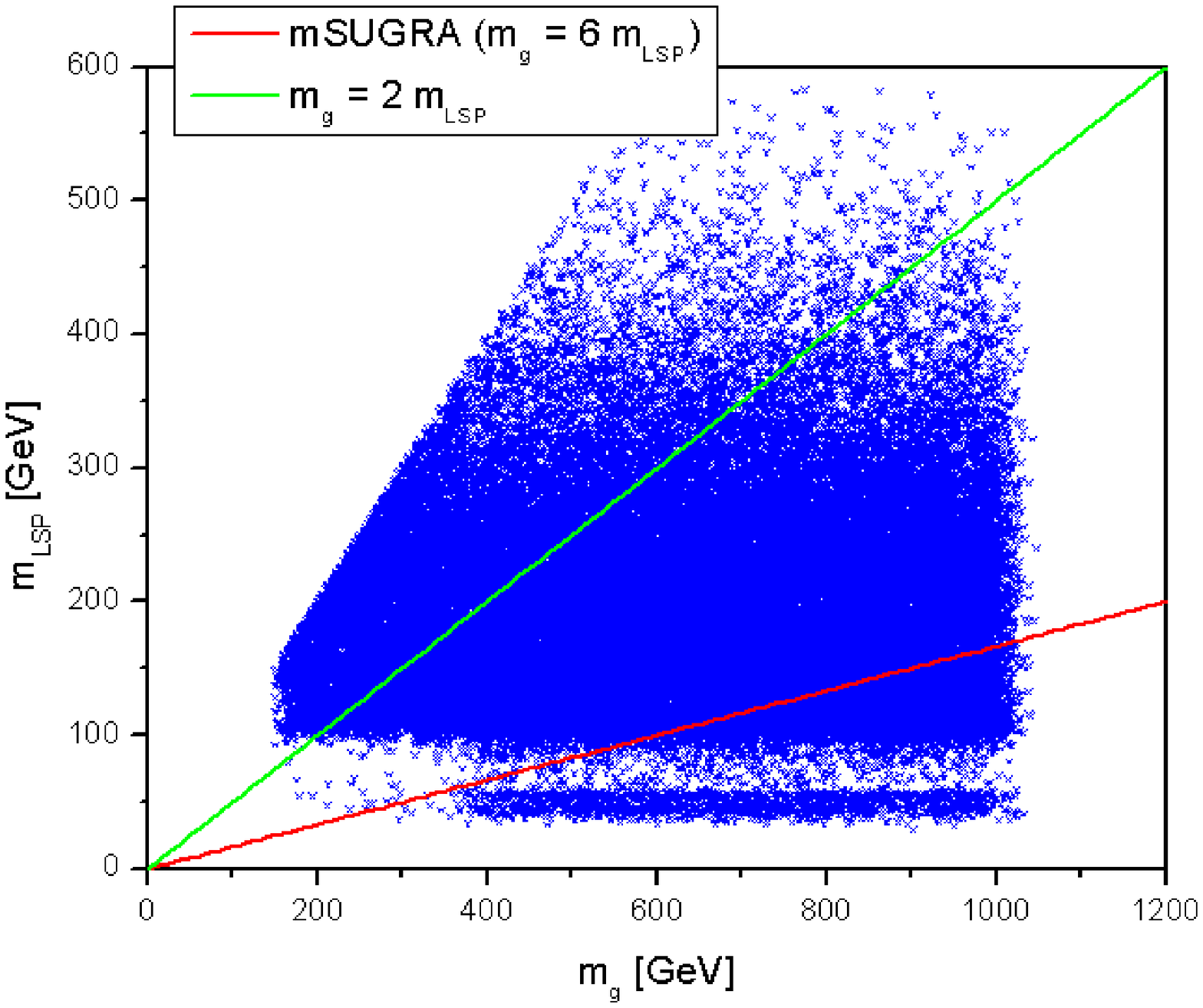}}
  \caption{For flat priors: (left)Light Higgs mass distribution
and (right)the correlation of the gluino and the LSP masses. The red line 
is the mSUGRA prediction.}
\label{fig1}
\end{figure}

In Fig.1 we see the predicted distribution of possible Higgs masses for the 
surviving models. Note that rather light masses are allowed relative to the 
$\sim 114$ GeV SM LEP search bound due to both reduced $Zb\bar b$ 
couplings as well as the existence of the $h\to 2\chi_1^0$ decay channel in 
some cases. This figure also shows a comparison of the gluino and LSP masses. 
In mSUGRA this mass ratio is $\sim 6$ but here all values are possible; the 
chunk removed from the lower left part of the figure is due to direct Tevatron 
searches. Note that rather light gluinos(with masses below 200 GeV) are 
possible when the gluino-LSP mass splitting is small since in this case the 
resulting jets are too soft to pass the Tevatron search cuts. These sparticles 
will be difficult to find at the LHC for similar reasons.

Fig.2 shows the predicted mass distributions for neutralinos and charginos; 
the LSP mass peak is near $\sim 150$ GeV. Note that many models predict 
charginos and possibly second neutralinos with masses not far above that of 
the LSP. This is due to the rather frequent occurance of either wino- or 
higgsino-dominated LSPs in our final model sample(s). Fig.3 shows the 
predicted mass splitting between the nLSP and the LSP as a function of the 
LSP mass. The range is quite extraordinary from over 500 GeV to below 1 MeV. 
The chunk removed from the plot on the lower left side is due to Tevatron 
stable particle searches as applied to charginos, \ie, for mass splittings 
below $\sim 100$ MeV these charginos are detector stable. This figure 
indicates that SUSY particle spectra with small mass splittings should be a rather 
likely possibility at the LHC. This would imply that stable particle searches will be 
important as will searches for sparticles decaying after traveling some finite 
distance inside the detector. 
Also shown in Fig.3 is the identity of the nLSP itself. 
While much of the time the nLSP is either the lightest 
chargino or the second neutralino, the 11 other possibilities can also occur 
with a reasonable frequency especially in the case of bino-like LSPs. 
Many of these scenarios are unusual and have not been well 
studied (if at all) at the LHC and may yield difficult or unusual signatures. 

\begin{figure}
\centerline{
  \includegraphics[height=.27\textheight]{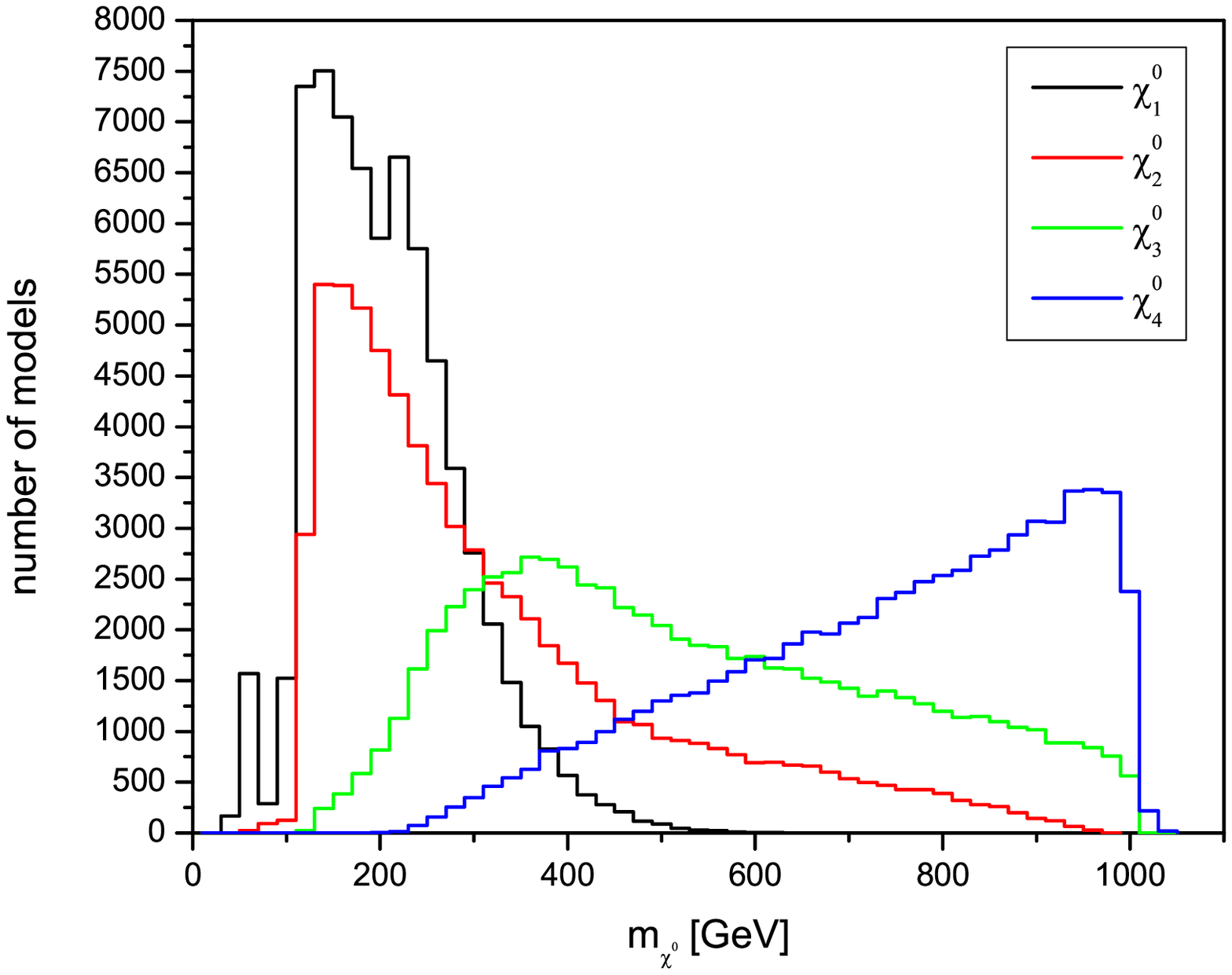}
\hspace*{-0.1cm}
  \includegraphics[height=.27\textheight]{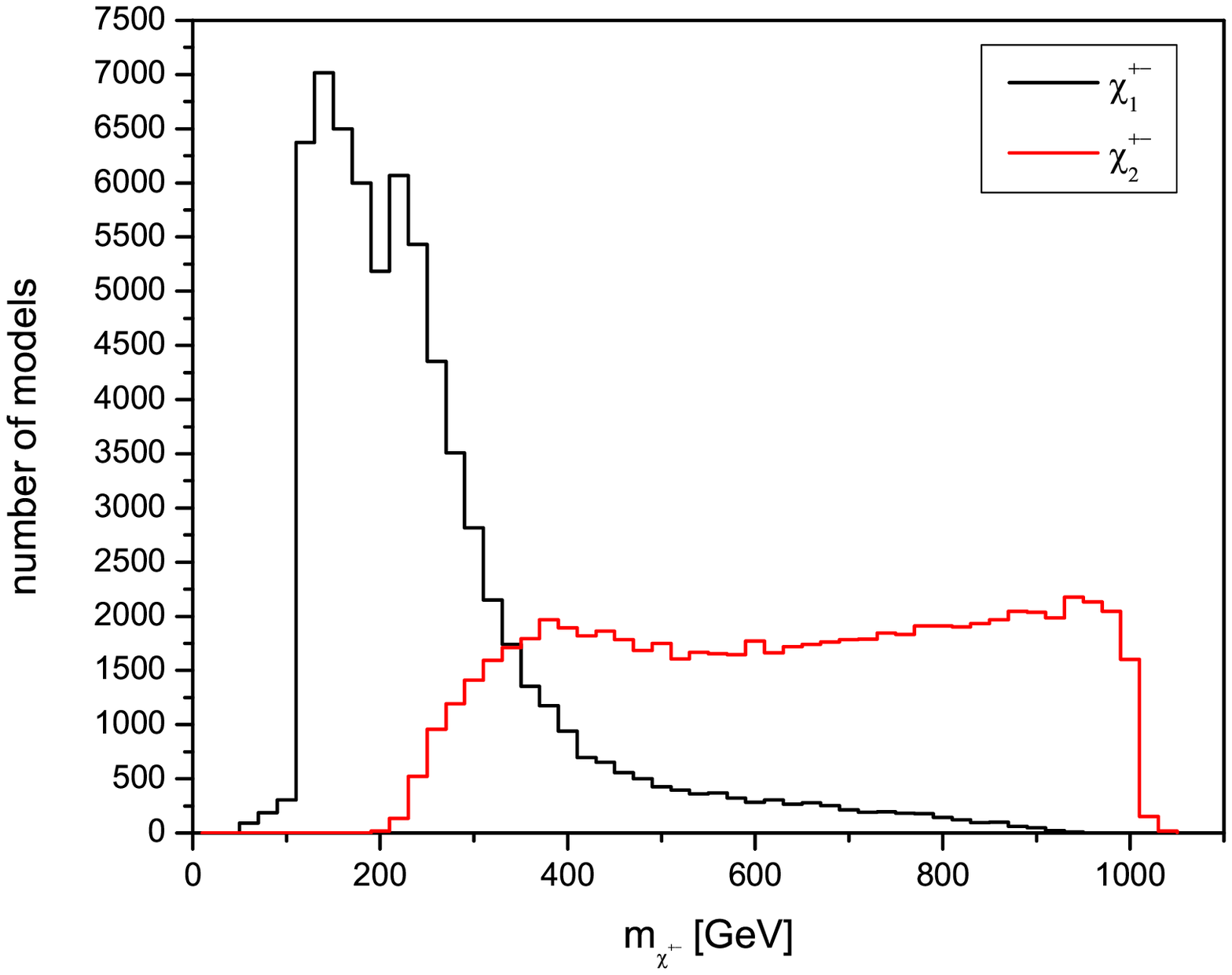}}
  \caption{Mass distributions for (left) neutralinos and (right)charginos 
assuming flat priors.}
\label{fig2}
\end{figure}

\begin{figure}
\centerline{
  \includegraphics[height=.27\textheight]{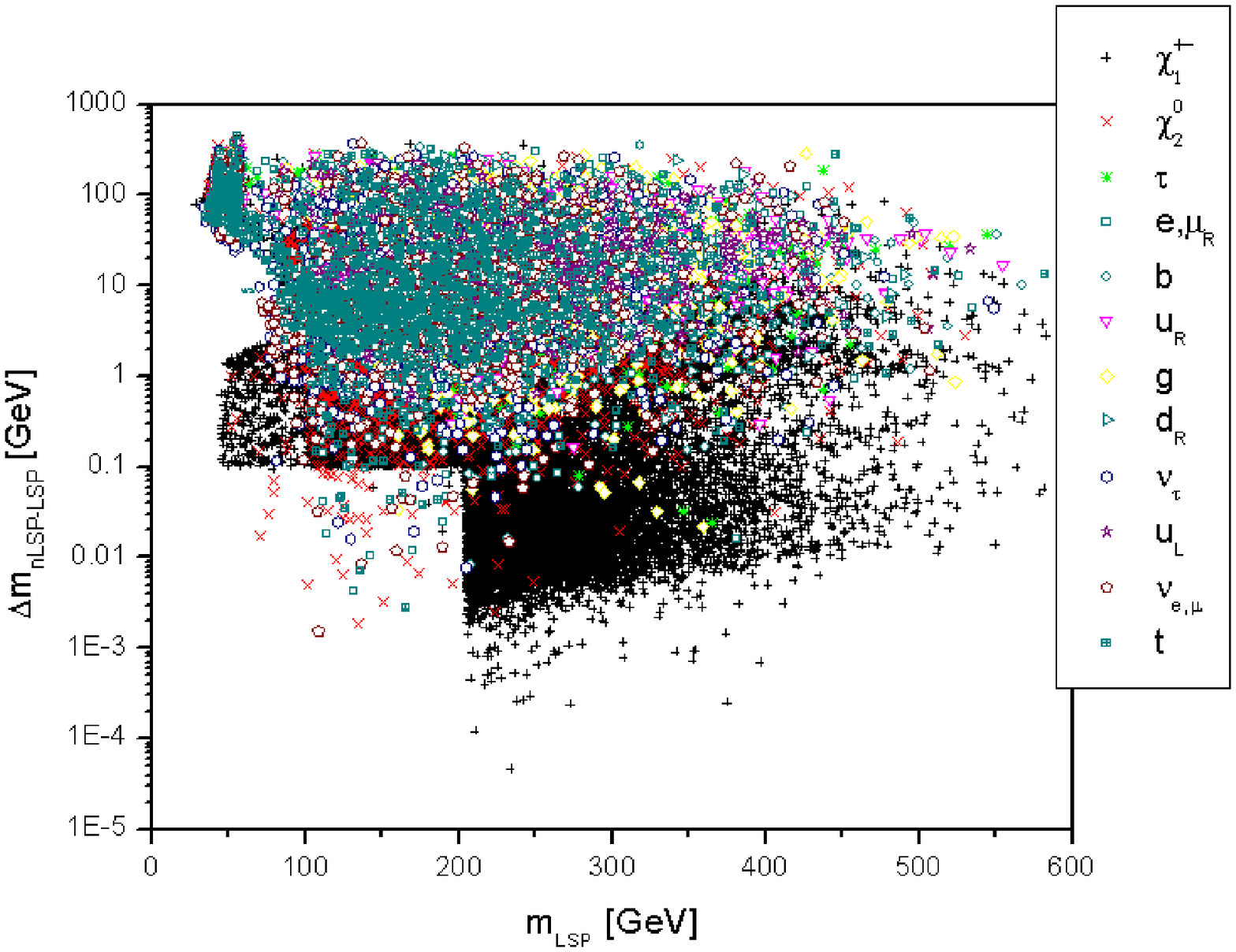}
\hspace*{-0.1cm}
  \includegraphics[height=.27\textheight]{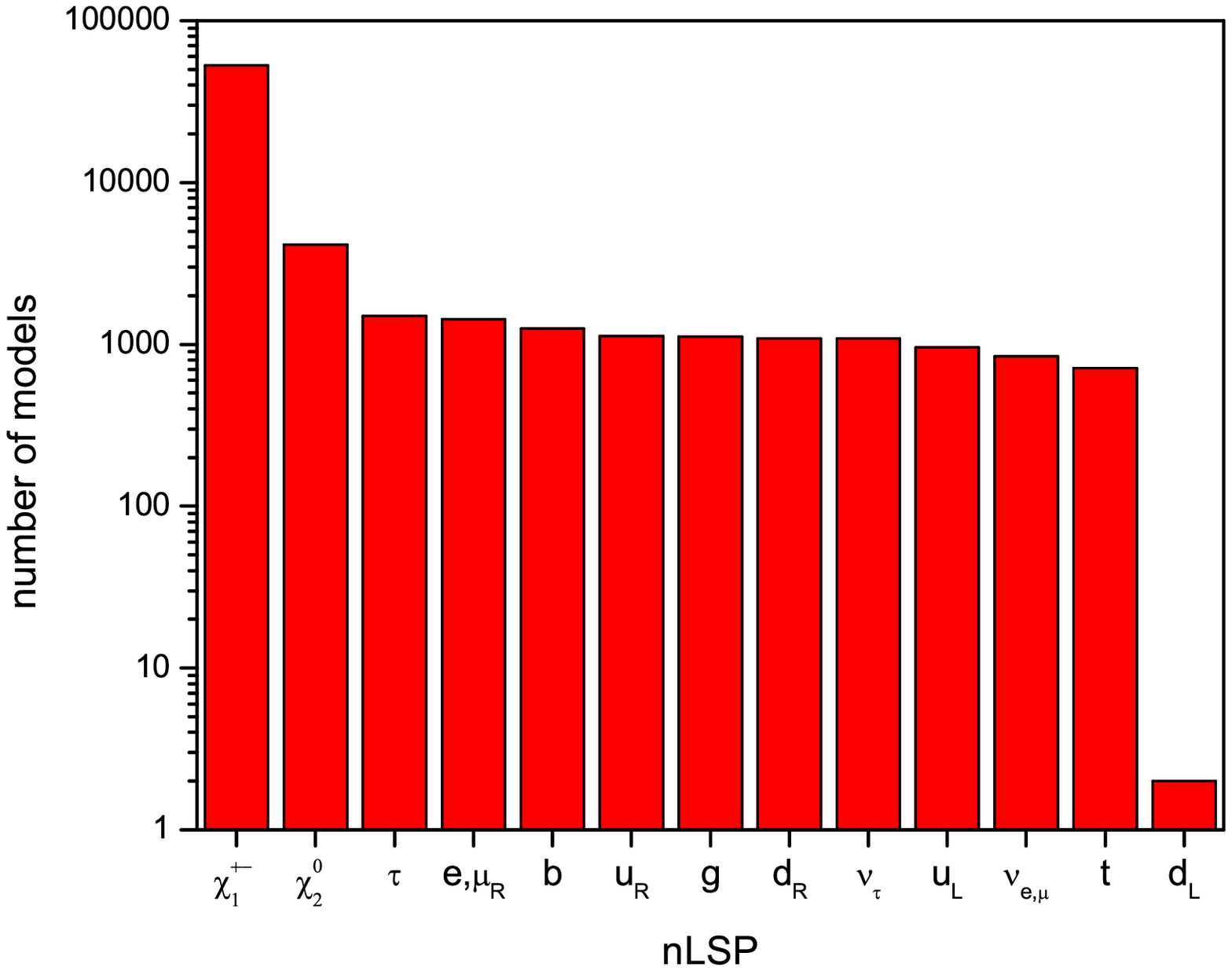}}
  \caption{(Left)The nLSP-LSP mass splitting as a function of the LSP mass 
and (right)the identity of the nLSP both for flat priors.}
\label{fig3}
\end{figure}

Fig.4 shows the LSP decomposition as a function of the nLSP-LSP mass 
splitting. Much of the time the LSP is relatively close to being a weak eigenstate. 
Here we see that the region of very small mass splittings is 
dominated by wino-like LSPs ($\sim 23.3\%$ of the time) while for large 
mass splittings the LSP is more likely to be bino-like ($\sim 17.1\%$ of the 
time). In the intermediate mass region the LSP is found to be mostly 
Higgsino-like ($\sim 45.6\%$ of the time).  Fig.4 also shows 
the distribution of predicted masses for the 
first and second generation squarks. As in the case of gluinos, light squarks 
remain a distinct possibility despite the Tevatron direct searches. Most 
commonly, this is again due to the potential small mass splitting that can 
occur between the squark and the LSP which leads to rather soft jets in the 
final state. Light squarks with such small mass splittings will also be difficult 
to see at the LHC.

\begin{figure}
\centerline{
  \includegraphics[height=.27\textheight]{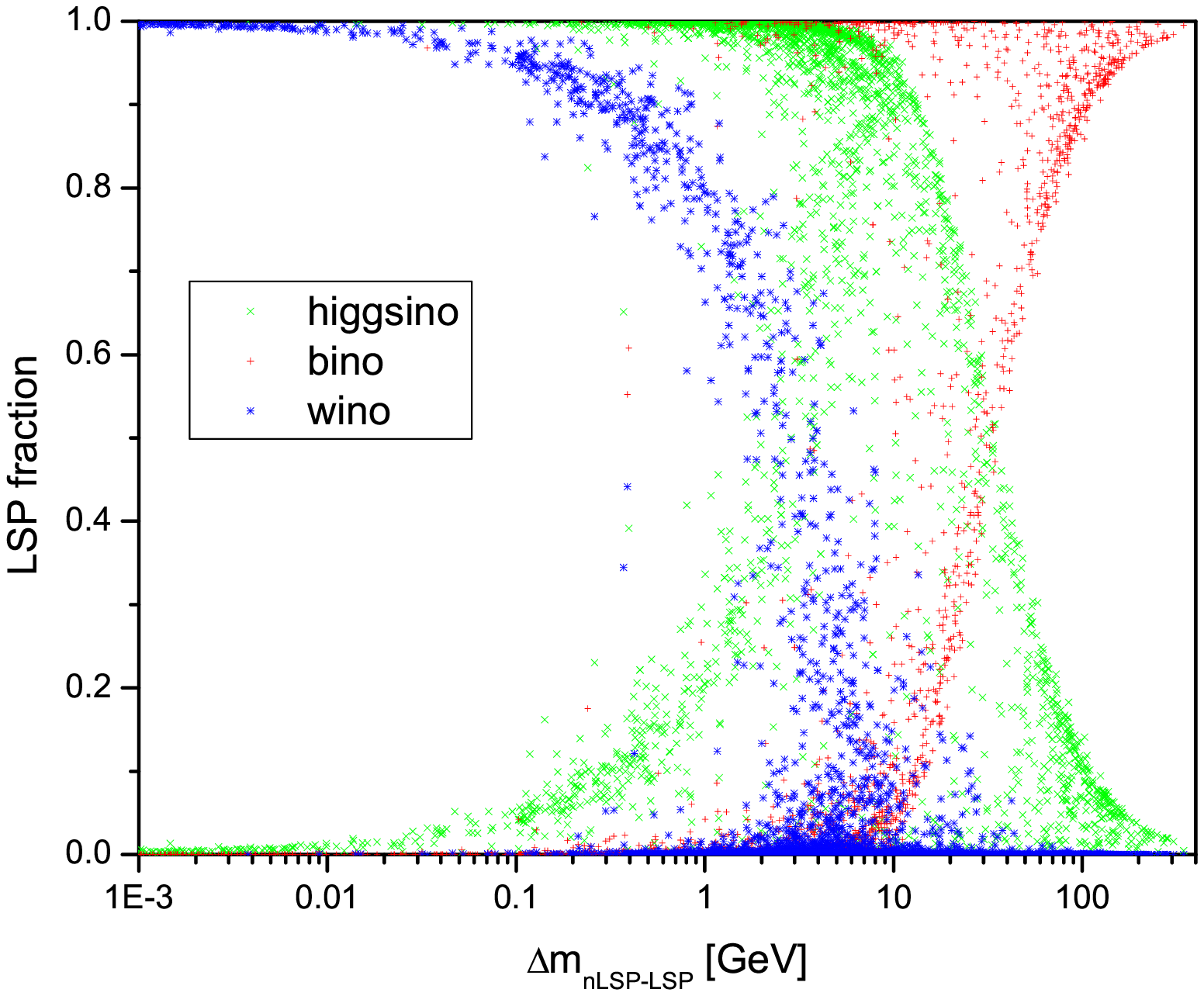}
\hspace*{-0.1cm}
  \includegraphics[height=.27\textheight]{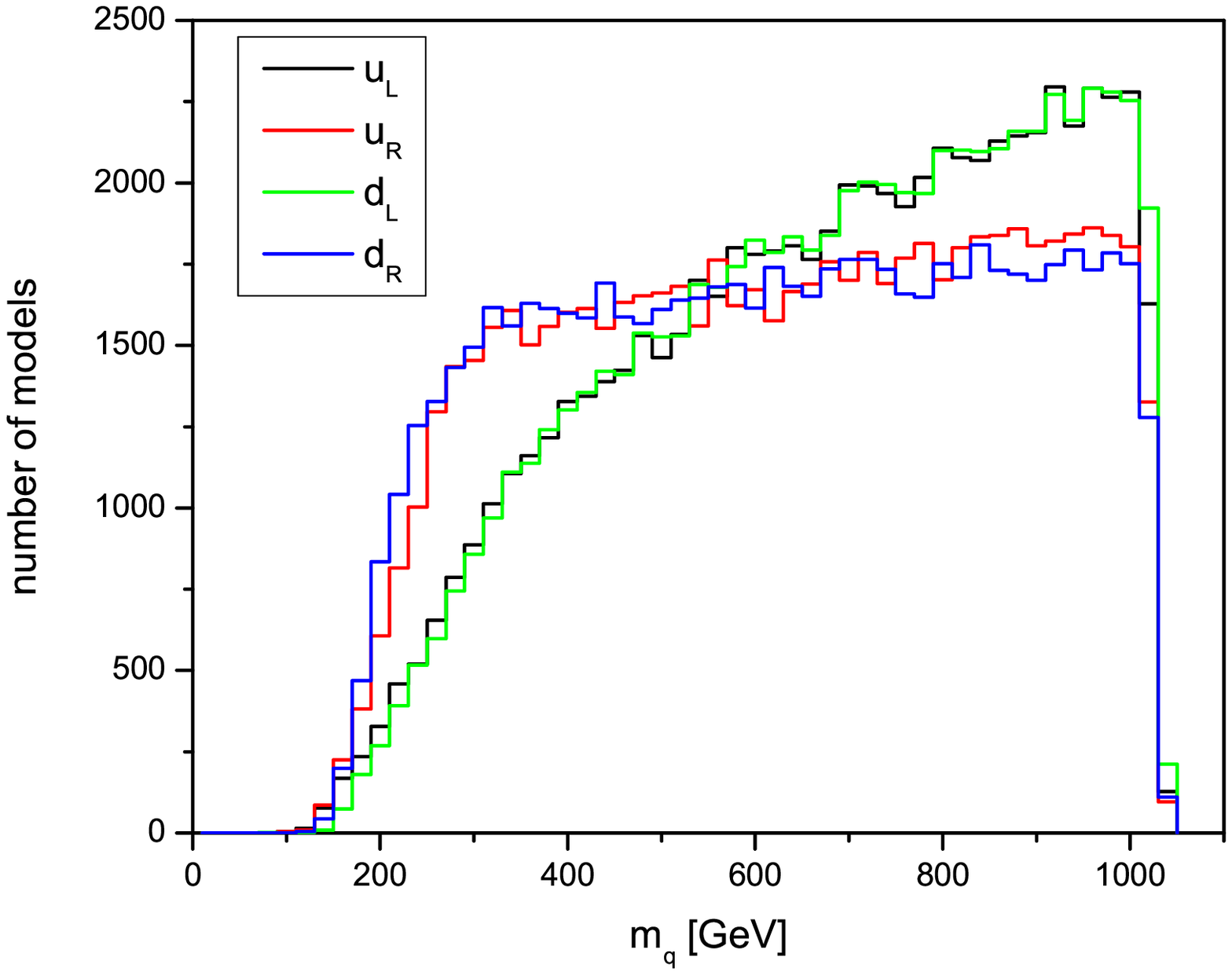}}
  \caption{(Left)The LSP composition for log prior models as a function of 
the nLSP-LSP mass splitting and (right) the predicted mass distribution for 
squarks for the flat prior case.} 
\label{fig4}
\end{figure}

As discussed above and in {\cite {us,talks}} the predictions of the MSSM can be vastly 
different from what one finds in any of the well-studied SUSY breaking scenarios. This 
can lead to unusual signatures at the LHC which may even be missed by conventional searches.


\begin{theacknowledgments}
Work supported in part by the Department of Energy, Contract DE-AC02-76SF00515.
\end{theacknowledgments}

\end{document}